\documentclass[twocolumn,showpacs,prb,amsmath,amssymb,floatfix*]{revtex4}
\usepackage{graphicx}
\usepackage{dcolumn}
\begin{document}

\preprint{Nanotubes dispersed in liquid crystal}

\title{Isotropic to Nematic Phase Transition in Carbon Nanotube dispersed Liquid Crystal Composites}

\author{Rajratan Basu, Krishna P. Sigdel and Germano S. Iannacchione\footnote{electronic address: gsiannac@wpi.edu}}
\affiliation{Order-disorder phenomena laboratory, Department of
Physics, Worcester Polytechnic Institute, Worcester, Massachusetts
01609, USA}

\date{\today}


\begin{abstract}

A high-resolution dielectric and calorimetric study of the
isotropic (I) to nematic (N) phase transition of carbon nanotube
(CNT) dispersed liquid crystal (LC) functional composites as a
function of CNT concentration is reported. The evolution of the
\emph{I-N} phase transition, the temperature dependence of local
nematic ordering formed by dispersed CNTs in the LC media and the
transition enthalpy were coherently monitored. Anisotropic CNTs
induce local deformation to the nematic director of LC and form
lyotropic \emph{pseudo-nematic} phase in the LC media. Results
clearly indicate the dramatic impact of dispersed CNTs on both the
isotropic and nematic phases of the composite.

\end{abstract}


\maketitle


Carbon nanotubes (CNTs) dispersed in a nematic liquid crystal
(LC)represent a  versatile functional composite that has gained
interest in recent years for inducing parallel alignment of CNTs,
improving electro-optic effect and switching behavior of LCs
 \cite{Michael02,Dierking1,Dierking2,Baik,Kamat,Basu1}. The LC+CNTs system is a unique
assemblage of an anisotropic dispersion (CNTs) in an anisotropic
media (LC), which makes it an attractive physical system to study
the phase transition phenomena. At the isotropic (I) to nematic
(N) transition, the orientational  order can  be described by a
symmetric and traceless second rank tensor ($Q_{ij}$)  which  can
be described by a scalar parameter $S(T)$ on short length scale
and on a longer length  scale by a vector $\hat{n}$ called nematic
director \cite{deGennes93}. In particular, orientational coupling
of LC and MWCNTs has an impact on both the nematic and isotropic
phases of LC as well as on \emph{I-N} transition. These can be
probed by studying the dielectric behavior of the composite.
Dielectric constant of an anisotropic material like LC is
orientation dependent. A nematic LC confined between parallel
plate electrodes maintains a constant director, $\hat{n}$, due to
plate boundary conditions. Planar LC molecules, being
perpendicular to the probing field, show smallest dielectric
constant, $\varepsilon_{\bot}$; and homeotropically oriented
(parallel to the probing field) LC molecules exhibit highest
dielectric constant, $\varepsilon_{\|}$, assuming the LC is
positive dielectric anisotropic. The dielectric anisotropy,
$\Delta\varepsilon$ = ($\varepsilon_{\|}$ - $\varepsilon_{\bot}$),
of LCs is one of the most important features to study their phase
transitions. The average dielectric constant in a complete
isotropic mixture is given by $\bar{\varepsilon}$ =
($\varepsilon_{\|}$ + 2$\varepsilon_{\bot}$)/3 =
$\varepsilon_{iso}$. In the uniaxial nematic phase the average
dielectric constant can be written as $\bar{\varepsilon}$ =
(\emph{a} $\varepsilon_{\|}$ + \emph{b} $\varepsilon_{\bot}$)
which is lower than the extrapolated value of $\varepsilon_{iso}$.
$\Delta\varepsilon$ depends on temperature and is proportional to
the scalar order parameter, S(\emph{T}). Now, as the system
reaches the complete disorder, \emph{i.e}., isotropic liquid, the
order parameter drops down to zero. This leads to the idea of
having no temperature dependence of $\varepsilon_{iso}$.

\begin{figure}
\includegraphics[scale=0.40]{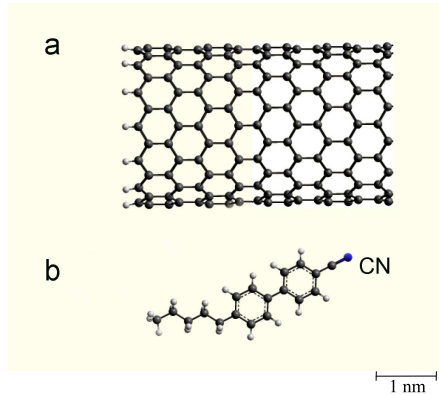}
\caption{ \label{Figure1}Cartoons of \textbf{a)} one carbon
nanotube structure, \textbf{b)} one 5CB liquid crystal molecule.
The diagram of the MWCNT and 5CB are shown approximately to scale.
Black spheres are carbon atoms and white spheres are hydrogen
atoms.}
\end{figure}

In this paper, we present an experimental study of the nematic to
isotropic phase transition in an LC+CNTs system for six different
concentrations of multiwall carbon nanotubes (MWCNTs) in LC media,
such as 0.05, 0.1, 0.15, 0.2, 0.25, and 0.3 wt $\% $. Average
dielectric constant ($\bar\varepsilon$) of such a system reveals
information about local as well as long range nematic ordering.
Calorimetric studies probe energy fluctuations of the phase
transition character. Thus, a combined \emph{T}-dependent
dielectric and calorimetric investigation has been undertaken to
study the \emph{I-N} phase transition in LC+CNTs. This work
reveals compelling evidence that nanotube aggregations create
local and isolated short range orientation orders in the LC media
which become prominent in the isotropic phase but does not cost
any extra energy fluctuations for low CNT concentration.

For these experiments, the well characterized liquid crystal
4-Cyano-4'-pentylbiphenyl (5CB) has been used as anisotropic host
for MWCNTs. Pure 5CB (2 nm long and 0.5 nm wide, weakly polar
molecule with M = 249.359 g/mol) has a weakly first-order
isotropic to nematic phase transition at $T^{0}_{IN} = 308K$ and
strongly first order nematic to crystal transition at $
T^{0}_{cr-N} = 295.5K $. Six different concentrations of MWCNT
sample (containing nanotubes 5 - 30 nm in diameter and 1-5 $\mu$m
in length) in 5CB were prepared. After mixing each wt $\%$ of
MWCNT sample with LC, the mixture was ultrasonicated for 5 hours.
In general, nanotubes aggregate due to attractive mutual Van der
Waals interactions but gentle ultrasonic agitations over a long
period reduce tendency of bundling of nanotubes and effectively
disperse and suspend them uniformly in a nematic LC matrix. Soon
after ultrasonication, each mixture was degassed under vacuum at
$40^{o}$C for at least two hours. In the LC+CNT system, helical
surface anchoring of LC molecules to the CNT-wall enhances
$\pi$-$\pi$ stacking by maximizing the hexagon-hexagon
interactions between this two species \cite{Park}. See Fig. 1. Due
to these interactions, self-assembled LC molecules impose
alignment on suspended carbon nanotubes along the nematic director
\cite{Michael02,Dierking1}.

\begin{figure}
\includegraphics[scale=0.4]{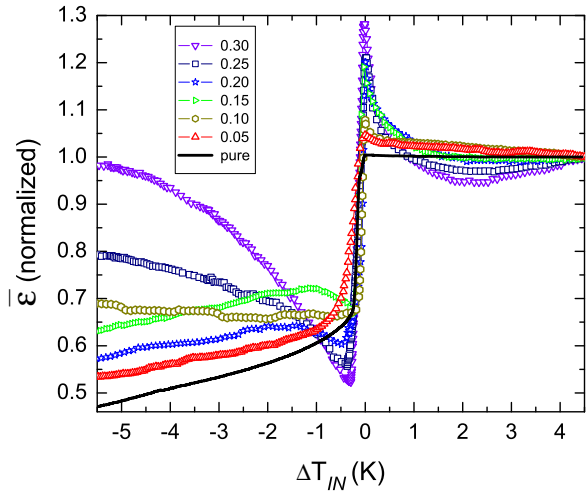}
\caption{ \label{Figure2}Normalized average dielectric constant
$\bar{\varepsilon}$ for pure 5CB and 5CB+MWCNTs as a function of
temperature shift, $\Delta T_{IN}$. The legend shows the
concentrations of dispersed MWCNTs in weight $\% $ in 5CB. The
absolutes values of $T_{IN}$ for all concentrations are shown in
Fig. 4a.}
\end{figure}

AC capacitance bridge technique \cite{Pilla,Bera,Foote} has been
used to measure $\bar\varepsilon$ as a function of temperature. A
droplet of each mixture was sandwiched between parallel-plate
capacitor configuration, 1 cm diameter and 100 $\mu$m thick,
housed in a temperature controlled bath. Dielectric measurements
were performed at very low probing field (5kV/m) and at 100 kHz
frequency. Comparison between the empty and sample filled
capacitor allows for an absolute measurement of
$\bar{\varepsilon}$(\emph{T}). Since low probing field strength
does not disturb the director orientation, no director
reorientation occurred. The samples that were used in dielectric
study were used in this AC calorimetry technique as well. The
sample cell for calorimeter consists of an aluminum envelope  of
length $\sim15~mm$, width $\sim8~mm$ and thickness $\sim0.5~mm$
with the three sides glued with super glue and was made using a
cleaned sheet of aluminum. The aluminum was cleaned with water,
ethanol and acetone using ultrasonic bath. Once the cell was dried
thoroughly the desired amount of sample LC+CNTs was loaded into
the cell and then a $120 \Omega$ strain gauge heater and $1
M\Omega$ carbon-flake thermistor were attached on its surfaces.
The filled cell then was mounted in the high resolution AC
calorimeter \cite{Paul68,Yao98} to study the phase transition
behavior. In the ac mode oscillating power $P_{ac}\exp(i\omega t)$
is applied to the cell resulting in the temperature oscillations
with amplitude $ T_{ac}$ and a relative phase shift between
$T_{ac}$ and input power, $\varphi=\Phi+\frac{\pi}{2}$ where
$\Phi$ is the absolute phase shift between $T_{ac}$ and the input
power. $\varphi$  also provides the information regarding the
order of the phase transition. With the definition of complex heat
capacity, $C^\ast = \frac{P_{ac}}{\omega T_{ac}}$, the heat
capacity at a heating frequency $\omega$ can be obtained
\cite{Germano98,Roshi04}. All calorimetric data presented here was
taken at a heating frequency of $31.25~mHz$ at a scanning rate of
$1~ K~h^{-1}$. For all LC+CNTs samples each heating scan was
followed by a cooling scan and experienced the same heating
history.

\begin{figure}
\includegraphics[scale=0.4]{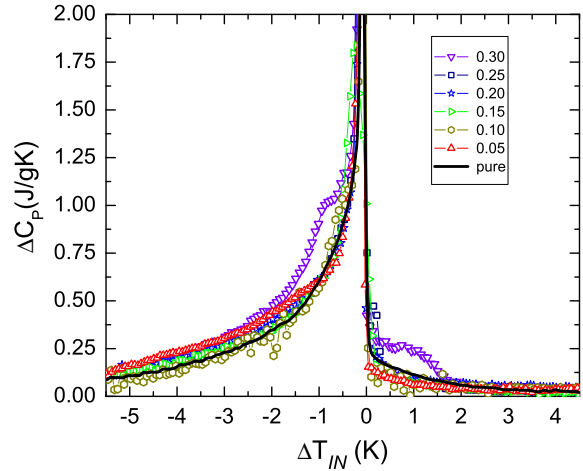}
\caption{ \label{Figure3} Excess specific heat $\Delta C_{p}$ for
pure 5CB and 5CB+MWCNTs as a function of temperature shift,
$\Delta T_{IN}$. The legend shows the concentrations of dispersed
MWCNTs in weight $\% $ in 5CB. The absolutes values of $T_{IN}$
for all concentrations are shown in Fig. 4a.}
\end{figure}

The normalized $\bar{\varepsilon}$ for different concentrations of
LC+CNTs are shown in Fig. 2 as a function of temperature shift
$\Delta T_{IN}$. The temperature shift is defined as $\Delta
T_{IN} = T - T_{IN}$, where $T_{IN}$ is the \emph{IN} transition
temperature for each concentration. The transition temperature is
defined as the temperature where $\bar{\varepsilon}$ shows the
first discontinuity while entering the $\emph{N+I}$ phase
coexistence region from isotropic phase and was determined from
$\bar{\varepsilon}$ vs. $T$ curves. Due to their high aspect
ratio, CNTs also exhibit dielectric anisotropy and the value of
average dielectric constant of aggregated CNTs is much larger than
that of LCs \cite{basu}. The addition of a very tiny amount of
MWCNT sample causes large increment in dielectric constant for the
LC+CNTs composites. To compare the dielectric behaviors properly
for all the concentrations, the dielectric constants are
normalized to the highest temperature (315 K) point studied. Bulk
5CB exhibits the classic temperature dependence of the dielectric
constant, showing nematic to isotropic phase transition at $T_{IN}
= 308.1 K$, seen in Fig. 2 and Fig. 4a. Above the transition
temperature the dielectric constant flattens out in the isotropic
phase and shows no temperature dependence at all
($\partial\varepsilon_{iso}/\partial T \neq \emph{f}(\emph{T})$),
indicating that, bulk 5CB reaches complete disorder state having
order parameter, \emph{S(T)} = 0. \emph{I-N} phase transition for
pure 5CB has been found at $ T_{IN} = 307.8~K$ by using the AC
calorimetry technique. The excess heat capacity, $\Delta C_{p}$,
associated with the phase transition can be determined by
subtracting an appropriate background from total heat capacity
over a wide temperature range \cite{Germano98,Germano04}. The
resulting $\Delta C_{p}$ data for LC+CNTs samples studied are
shown in Fig. 3 over a $10K$ temperature range window about
$\Delta T_{IN}$. The transition temperature is defined as the
temperature of entering the $\emph{N+I}$ phase coexistence region
from isotropic phase and was determined from $C_{p}^{"}$ vs. $T$
curves. The excess heat capacity peaks are being shifted farther
to zero of the temperature axis as the CNTs concentration
increases, but the shifting does not have a fixed trend as also
observed in dielectric results. The small shift on transition
temperatures due to presence of CNTs is because of the large
density differences between the liquid crystal and CNTs
\cite{Paul08}. The $\emph{I-N}$ transitions evolve in character.
The nature of $\Delta C_{p}$ wings in Fig. 3 for \emph{I-N}
transitions are the same for all samples. The highest
concentration (0.3 wt $\% $) studied shows two extra features at
both sides of the transition. This is possibly due to the
transition of \emph{pseudo-nematic} phases \cite{Basu1} of
dispersed CNTs in LC media.

\begin{figure}
\includegraphics[scale=0.50]{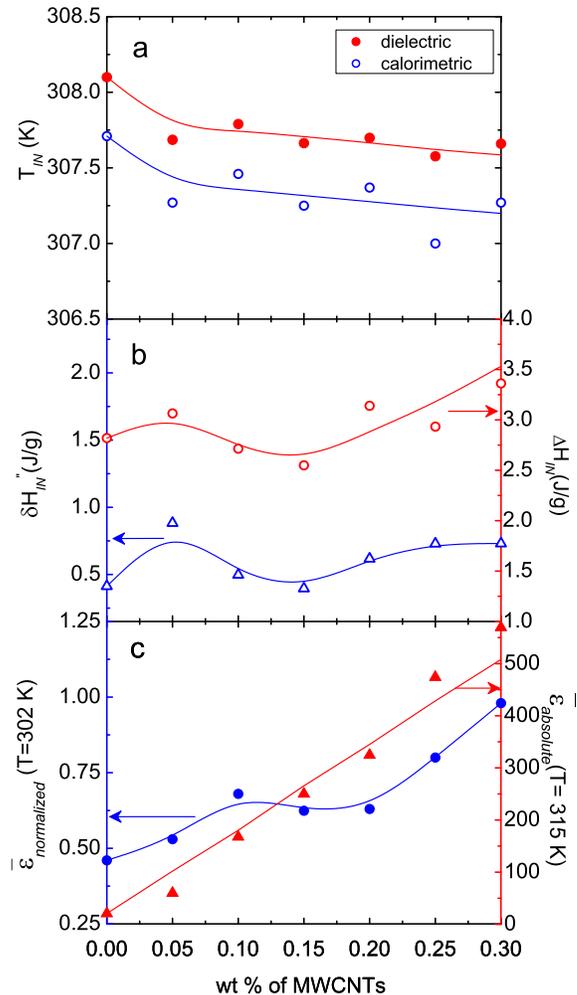}
\caption{ \label{Figure4} \textbf{a)} The transition temperature,
$\Delta T_{IN}$, as a function of MWCNTs concentration and
comparison between the two techniques used for the measurements,
\textbf{b)} AC- enthalpy (right) and imaginary enthalpy (left) as
a function of MWCNTs concentration, for details see text,
\textbf{c)} The normalized dielectric constant,
$\bar{\varepsilon}$, as a function of MWCNTs concentration at the
starting temperature, 302 K (left) and absolute dielectric
constant in the deep isotropic phase (315 K) as a function of
MWCNTs concentration (right).}
\end{figure}

Different weight concentrations of the 5CB+MWCNTs functional
composites reveal a dramatic change in the dielectric behavior as
well as in nematic to isotropic phase transition behaviors.
Smaller amount of CNTs dispersed and suspended in nematic LC media
is not frozen in the system; rather they diffuse in the nematic
matrix due to the thermal fluctuation. This can be visualized as
annealed random variables evolving with time. The nano dynamics of
CNTs induce local deformation of LC nematic director. The higher
the CNT concentration the larger local deformation occurs.
Dielectric behaviors also provide information about the molecular
arrangements in a particular system and the larger the
$\Delta\varepsilon$ the smaller electric field is needed to make
an anisotropic system respond to it. $\Delta\varepsilon$ of
aggregated CNTs is much higher than that of LCs.
$\Delta\varepsilon$ increases locally in the LC+CNTs system in
addition of CNTs. This makes the locally ordered domains, due to
the presence of CNTs, more responsive to the probing field, hence
the dramatic evolution in the nematic phase with different CNTs
concentrations, clearly seen in Fig. 2. Fig. 4c shows the
normalized $\bar{\varepsilon}$ at the starting temperature (302 K)
as a function of concentration of CNTs. Clearly there is a
crossover region from 0.1 to 0.2 wt $\%$. Above the crossover
region the locally ordered domains tend to align along the field
and below that crossover region, the system is not that responsive
to the low probing field. This explains the downward curvatures in
the nematic phase for the higher concentrations observed in Fig.
2. Due to the elastic coupling and strong anchoring of LCs to
nanotube-surfaces, both the species co-operatively create local
short range orientation order. In addition to that, CNTs
themselves form lyotropic nematic phase when they are dispersed in
a fluid \cite{Song}. The curvatures in normalized
$\bar{\varepsilon}$ in the isotropic phase for LC+CNTs seen in
Fig.2 are the evidence of presence of local \emph{pseudo-nematic}
order formed by CNTs in an isotropic liquid. The larger the CNT
concentration the more curvature in $\bar{\varepsilon}$ occurs
above $T_{IN}$ confirming that the formed local order is lyotropic
\emph{pseudo-nematic} phase. It has been observed that at about
315 K the $\bar{\varepsilon}$ becomes independent of temperature
for all concentrations, confirming that the thermal energy clears
out the \emph{pseudo-nematic} order and the system reaches
complete isotropic state. The total enthalpy change associated
with a first order phase transition is the sum of the
pre-transitional enthalpy and latent heat. Due to partial phase
conversion $\left(N\leftrightarrow I\right)$ during a $T_{ac}$
cycle, typical $\Delta C_{p}$ values obtained in two phase
coexistence region are artificially high and frequency dependent.
The procedure for calculating $\delta H$ and $\Delta H$ is given
in ref.~\cite{Germano98}. An effective enthalpy change $\Delta
H^{\ast}_{IN}$ which includes some of the latent heat contribution
can be obtained by integrating observed $\Delta C_{p}$ peak . Here
we performed a complete integration over the entire $\Delta C_{p}$
peak over a wide temperature range of around $300K$ to $312K$ for
all 5CB+CNTs samples  to get effective enthalpy change $\Delta
H^\ast_{IN}$ associated with $IN$ transition. The integration of
the imaginary part of heat capacity gives imaginary enthalpy,
$\delta H^{"}_{IN}$, which indicates the first order character of
the transition or the dispersion of energy in the sample. As fixed
frequency used for this work $\delta H^{"}_{IN}$ is only
approximately proportional to the transition latent heat even
though it is the measure of the dispersive part of the complex
enthalpy. The proportionality constant is different for various
CNT concentrations due to different values of the two phase
conversion rate. The results of $\Delta H^\ast_{IN}$ and $\delta
H^{"}_{IN}$ as a function of CNT concentrations for all 5CB+CNTs
samples are presented in Fig. 4b. Since $\Delta C{p}$ values for
the different concentration were different ac-enthalpy and
imaginary enthalpy are also influenced by the CNT concentrations.
$\Delta H^{\ast}_{IN}$ slightly increases with the increase in CNT
concentration. The imaginary-enthalpy, $\delta H^{"}_{IN}$
fluctuates slightly, but overall it also has the increasing trend
with increase in CNT concentration. In ac-calorimetric technique
the uncertainty in determining the enthalpy is typically 10$\%$.

As demonstrated above, both dielectric and calorimetric
measurements indicate that MWCNTs dispersed in 5CB create local
lyotropic \emph{pseudo-nematic} phase, but that does not change
the net fluctuation of energy much during \emph{I-N} transition.
Evolving wings of normalized $\bar{\varepsilon}$ in the nematic
phase in Fig. 2 indicate that local domains due to the presence of
CNTs become more responsive to the probing field as CNT
concentration increases in the system. A fixed trend in the
curvature of $\bar{\varepsilon}$ in isotropic phase with
increasing CNT concentration manifests the para-nematic phase
formed by dispersed CNTs in an isotropic fluid. Calorimetric study
conspicuously evidences that tiny amount of suspended CNTs in LC
does not cause considerable enthalpy fluctuation about the
\emph{I-N} transition region. However, in the highest
concentration (0.3 wt $\% $), the \emph{pseudo-nematic} phase
formed by CNTs is strong enough and releases considerable amount
of energy while melting, getting displayed in the $\Delta C_{p}$
vs. $\Delta T_{IN}$ graph. Additional electric field dependent
experimental studies are planed to fully understand the character
of field response of local domains in LC+CNTs system.


\newpage

\newpage

\bibliography{lc+cnt}

\end{document}